# Integrating Cascade Pumped Micro-Hydro Storage: A Sustainable Approach to Energy and Water Management

Oraib Dawaghreh[1], Sharaf K. Magableh[1], Xuesong Wang[1], Mohammad Adnan Magableh[2], Caisheng Wang[1]*

[1]Department of Electrical and Computer Engineering, Wayne State University, Detroit, United States

[2]Electrical Engineering Department, Jordan University of Science and Technology, Irbid, Jordan

oraib.dawaghreh@wayne.edu, sharaf.magableh@wayne.edu, xswang@wayne.edu, mohammad.magableh@ieee.org, cwang@wayne.edu

*Abstract*—As traditional large hydropower has been extensively exploited, micro-hydro systems have caught research increasing interest. New engineering challenges arise in developing micro-hydro systems in areas with significant elevation but prohibitive horizontal distances between primary reservoirs. This study addresses these challenges by proposing a cascade-pumped micro-hydro storage (CPMHS) system that leverages intermediate reservoirs to bridge long horizontal distances, enabling efficient energy transfer and storage. The methodology utilizes naturally occurring lakes with substantial head heights but limited feasibility for direct pumped storage due to horizontal separations. Integrating smaller, strategically placed intermediate reservoirs maximizes energy capture along the cascading path, making pumped storage viable in geographically constrained locations. The proposed system will enhance energy generation potential and provide additional benefits for water management. Using geographical data and a detailed case study focused on Mountain Lake and surrounding lakes, this paper demonstrates the energy efficiency and viability of cascade-based micro-hydro storage. A practical methodology for implementing CPMHS systems is proposed and validated by case studies. An optimization framework is developed for efficient energy capture in regions with challenging topography.

## I. Introduction

The need for sustainable energy has reached new heights as the world tends toward deep carbonization to cut off carbon emissions and shift to renewable power sources [1]. Hydropower, one of the earliest forms of clean energy, remains a key player in this journey. With most of the major hydropower locations already utilized, there has been an increasing focus on discovering new sites, including micro-hydro plants [2]. These smaller systems, though often requiring creative engineering, are promising for powering remote areas, enhancing grid flexibility, and supporting local economies. Cascade hydropower emerges as an effective option for sustainable energy production by leveraging a series of dams and reservoirs [3]. Pumped hydro storage (PHS) is one of the most efficient and scalable options for stabilizing the grid, helping integrate unpredictable sources like wind and solar [4]. One promising application of PHS is in cascade systems, where a series of reservoirs harness natural elevation changes to create a continuous, sustainable flow of energy.

Cascade hydropower, when paired with micro-hydro storage, offers a resilient solution capable of addressing spatial and logistical constraints that limit traditional hydropower facilities [5]. This paper proposes the concept of cascade-pumped micro-hydro storage (CPMHS) by utilizing natural lakes that have significant elevation differences and long horizontal distances. In a cascade-PHS system, reservoirs are placed at different elevations, allowing water to be pumped between them in multiple stages. Instead of using a single large upper and lower reservoir, this approach distributes the process across several levels. This method improves hydro-generation, particularly in landscapes where there are significant horizontal distances between natural water bodies or varied elevation changes. Hence, the motivation for this study comes from investigating several promising lakes in Michigan that cannot be used for classical hydropower systems or even traditional PHS, due to the large horizontal distances between these existing lakes. Those cascade hydropower reservoirs will generate electricity and offer multiple water management benefits. This method supports regional development, reduces greenhouse gas emissions, and offers a renewable alternative to fossil fuels, making it a vital contributor to the global shift towards cleaner energy [6]. Additionally, the multi-stage design allows for scalability, adapting to varying topographical conditions and energy demands.

## II. State-of-the-art Literature Review

Numerous studies in the existing literature have examined the use of cascade hydropower, but generally, they have overlooked the methodology for selecting the optimal number of reservoirs and CPMHS. The authors in [7] introduced a new multi-stage robust scheduling approach for cascade hydropower systems, aiming at ensuring the feasibility of scheduling solutions and enhancing the use of hydropower reserves. The method presents a key proposition that precisely establishes the relationship between hydropower output and the operational security of the electricity grid. In [8], the authors examined the availability of renewable energy resources (RERs) in Pakistan. The study introduces a micro-hydropower plant located on a specific canal in Khyber Pakhtunkhwa, Pakistan, with modeling and optimization performed using RETScreen software. They simulated a 107 kW micro-hydro system over a 20-year period. The study confirms the technical and economic viability of the micro-hydro system, achieving a projected $0.049 per kWh cost of

energy. The findings suggest that the project would achieve payback by the fourth operational year. The authors in [9], explored an optimal capacity configuration approach for hybrid systems that integrate cascade hydropower, photovoltaic (PV), and wind. In a case study of the Yellow River's upper reaches in Qinghai, China, they determined that the ideal PV-wind capacity configurations are 561.2 MW and 651.8 MW under 3 and 5-segment transmission modes, respectively. They also found that as the PV capacity increases, cumulative fluctuations in combined PV-wind output first decrease and then rise, with an optimal PV-to-wind ratio of 0.744 to 0.256. In [10], a framework was developed to optimize short-term operational strategies for a cascade hydro-PV hybrid system. The study introduces a methodology to characterize net load, using a daily average net load and fluctuation coefficient. A model was then created to define the system's optimal short-term operation, including a synchronous peak-shaving strategy for cascade hydropower stations. The study revealed that single-peak loads were maintained on clear days with runoff below 30% of rated flow, while double-peak loads were required during high runoff days or under adverse weather conditions. This study introduces a methodology to optimally utilize a CPMHS system, as illustrated in Fig. 1. This is inspired by Michigan's extensive network of natural lakes.

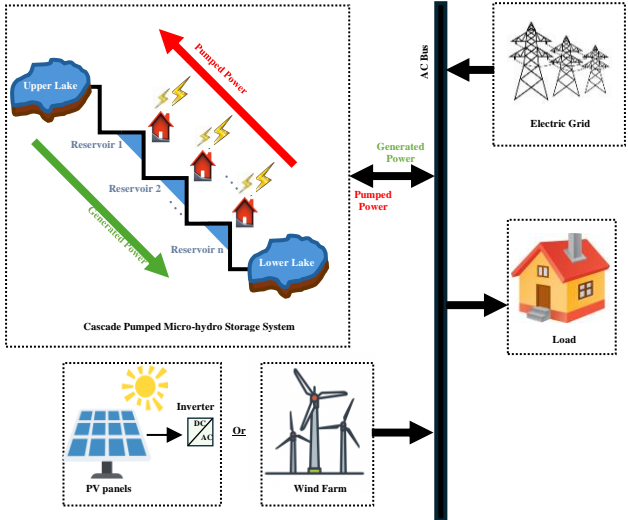

Fig. 1. Schematic of on-Grid Cascade Pumped Micro-Hydro Storage System.

Indeed, with over 11,000 inland lakes, Michigan offers vast, untapped potential for hydropower applications. Most of these lakes remain unexplored for traditional hydropower facilities or even as potential sites for PHS systems despite their significant elevation differences and wide horizontal distances. This abundance of natural resources presents an opportunity for sustainable solutions in renewable energy, which this study aims to address. Given the impracticality of a single PHS facility across large horizontal separations, the proposed approach determines the optimal number and placement of intermediate reservoirs to enable efficient energy storage and management. Hence, using advanced optimization algorithms in future designs, focused on maximizing power output based on real load demand data and geographic conditions, would be a powerful approach for leveraging these lakes. This strategy could greatly benefit authorities in Michigan and across the U.S., opening new pathways for efficient energy and water resource management. Fig.1 presents a schematic diagram of the grid-connected CPMHS system, illustrating the hybrid renewable energy components and CPMHS configuration designed to enhance regional energy and water resource management.

### III. PROPOSED MATHEMATICAL MODEL FOR CASCADE HYDROPOWER SYSTEM CASE DESIGN

Mathematical modeling provides a framework for simulating and optimizing system performance under specified constraints, ensuring efficient operation and resource utilization. This section presents the mathematical modeling approaches developed for both the CPMHS, setting the foundation for performance evaluation and system reliability. It is worth mentioning that other renewable generations like solar and wind will be used in the future work. This is to utilize the surplus power from those renewables in pumping the water throughout the cascaded reservoirs. The methodology introduced in this paper offers a case study focused on a hybrid renewable energy configuration. It also demonstrates the adaptability of the approach to integrate other renewable resources, such as solar or biomass, or combined applications, depending on the location and availability of renewable resources for specific designs and applications. Moreover, this model outlines a structured, validated methodology to guide effective workflow and operation within CPMHS systems.

#### A. Mathematical Modeling of CPMHS

CPMHS system functions similarly to batteries, storing energy in the form of potential energy for use when needed. Unlike traditional PHS systems, the CPMHS utilizes multiple interconnected reservoirs situated at varying elevations, allowing for energy generation and storage across several stages. This multi-layered approach distributes the energy process among several reservoirs, as illustrated in Fig. 2. The generated energy ($P_{\text{CPMHS}_{gen}}(t,i)$) and the pumping flow rate (outflow rate) ($q_p(t,i)$) between these stages can be modeled as shown in equations (1) and (2) [11].

$$P_{\text{CPMHS}_{gen}}(t,i) = \eta_t \times \rho \times g \times h \times q_t(t,i) \quad (1)$$

$$q_p(t,i) = \frac{\eta_p\, P_{\text{CPMHS}_{ch}}(t,i)}{\rho g h} \quad (2)$$

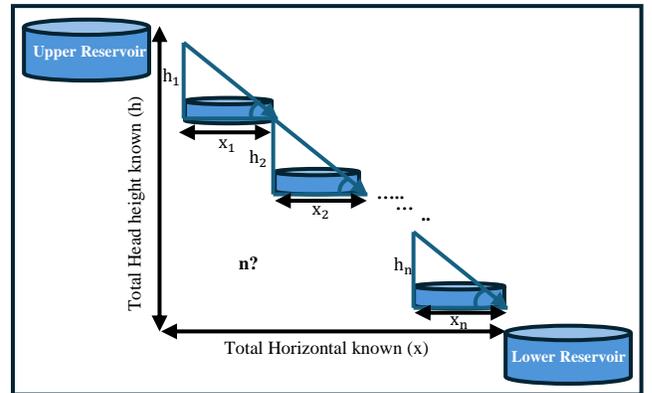

Fig. 2. Mathematical Representation of Stages in the CPMHS System.

In the equations, $\eta_t$, $\eta_p$ represent the efficiencies of the turbine-generator set and the pump, respectively. Additionally,

$\rho$ denotes the density of water, $g$ indicates gravitational acceleration, $h$ is the elevation difference, and $q_t(t,i)$ is the volumetric inflow rate of water in $m^3/s$ for reservoir $i$ at time $t$. For each reservoir $i$ at time $t$, the water level ($QOW(t,i)$) is influenced by the inflows, outflows, and losses due to evaporation and leakage ($\alpha$) as expressed in (3). Moreover, the storage capacity of each reservoir must remain within the specified minimum and maximum limits, as stated in (4).

$$QOW(t,i) = QOW(t,i-1)(1-\alpha) + q_p(t,i) - q_t(t,i) \quad (3)$$

$$QOW_{min}(t,i) \leq QOW(t,i) \leq QOW_{max}(t,i) \quad (4)$$

### B. Optimized Hybrid Power System Design: Flowcharts and System Configuration

Figs. 3, 4, and 5 illustrate a structured approach to optimizing a hybrid power system configuration, focusing on balancing energy generation, storage, and grid interactions to achieve an efficient and reliable system. Fig. 3 provides an overarching hybrid power system optimization process flowchart. Starting with the initialization of system configurations, it integrates essential input parameters and measured data, including hourly load profiles, wind turbine characteristics, costs, and operational constraints.

Through solar and/or wind power modeling, the system determines if the generated power meets demand and handles scenarios of both excess and deficit power. This step allows the system to decide whether to export power to the grid or store it in a pumped hydro system for future use. An iterative algorithm assesses and refines system configurations based on targeted fitness criteria, ultimately leading to the optimal solution.

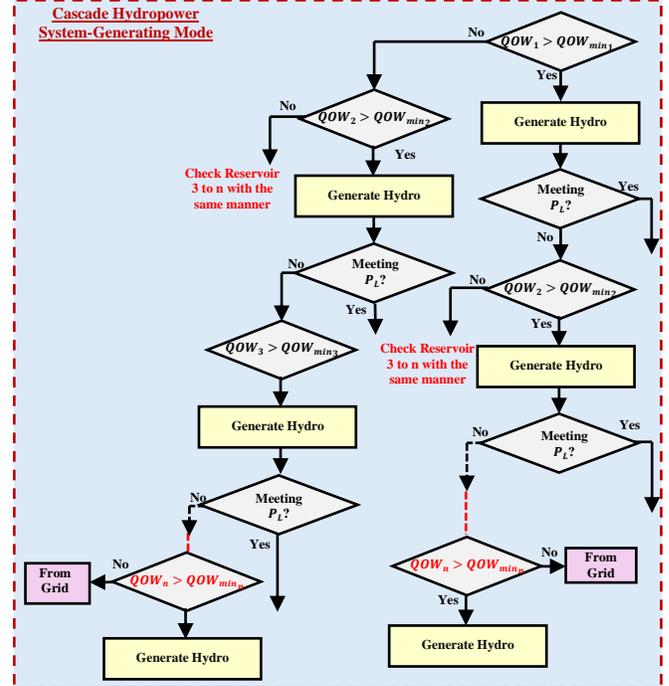

Fig. 4. Decision-Making Process for Efficient Power Generation Across Multiple Reservoirs Based on Water Availability and Demand.

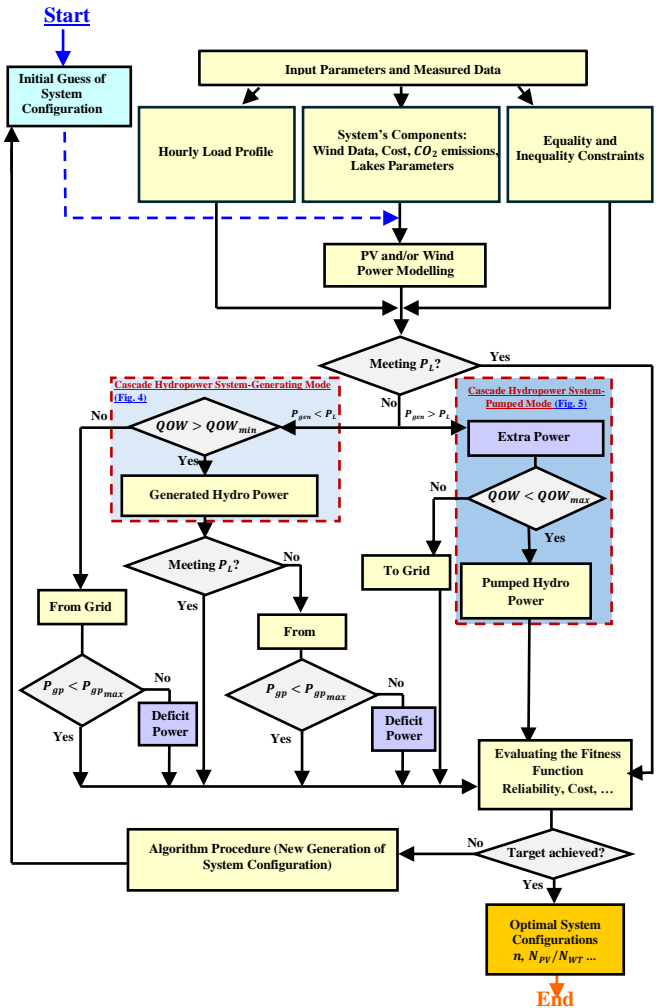

Fig. 3. Hybrid Power System Optimization Process: Evaluating and adjusting wind power and grid interactions to achieve the optimal configuration.

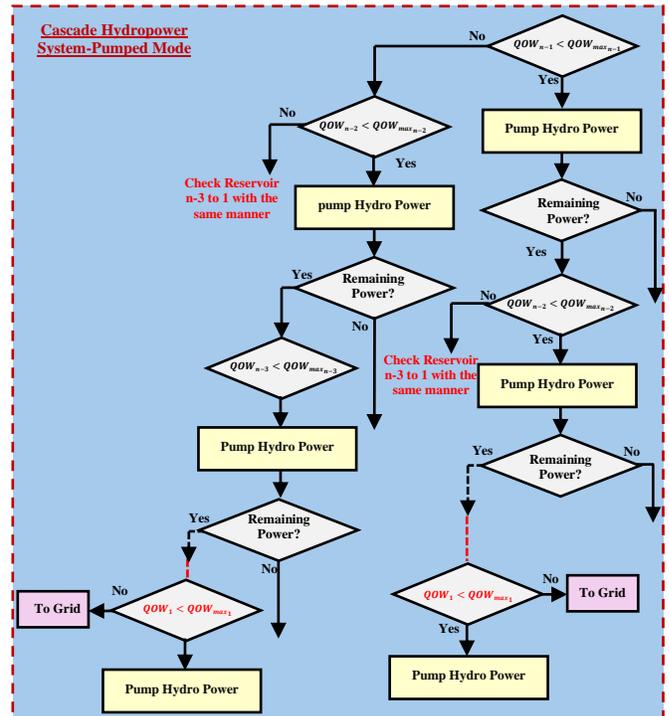

Fig. 5. Control Logic for Energy Storage by Pumping Water to Higher Reservoirs Based on Water Availability and Power Surplus.

Fig. 4 illustrates the cascade hydropower system in self-generating mode, where interconnected reservoirs coordinate to generate power efficiently based on water availability and demand. The flowchart outlines a series of decisions, starting with checks on upstream reservoir outflows ($Q_{out}$) to determine if they exceed the required thresholds for hydro generation. If sufficient water is available, the system initiates power generation at the appropriate stage; otherwise, it moves downstream to assess other reservoirs. When power demand exceeds hydro generation capabilities, the system evaluates whether energy should be supplemented from the grid. This dynamic process ensures continuous power generation by leveraging available water resources across multiple reservoirs while optimizing grid interactions to meet demand reliably.

Additionally, Fig. 5 zooms in on the cascade hydropower system's pumped mode, showcasing a detailed flowchart of managing hydroelectric power within cascading reservoirs. This subsystem assesses whether upstream reservoirs have excess water, enabling lower reservoirs to pump water when necessary. Decisions within this cascade system hinge on factors like water availability, flow rate, and storage capacity in each reservoir. When excess power is available, it triggers the pumping mechanism, efficiently storing energy by moving water to higher elevations.

This cascaded setup facilitates renewable energy storage and ensures responsive power flow between reservoirs and the grid, optimizing overall system performance. Note that all floating arrows indicating "yes" and "No", in Figs. 4 and 5, respectively, proceed toward evaluating the fitness function as Figs. 4 and 5, which are the zoom of the generating and the pumping modes of Fig. 3. Together, these figures outline a holistic and iterative process that considers both wind and hydropower sources, utilizing cascading reservoirs for enhanced energy storage and control. This approach supports the optimal use of renewable resources, enabling a robust and sustainable hybrid power system.

## IV. ANALYSIS OF CASE STUDY IMPLEMENTATION

This section presents a case study from Michigan, assessing the potential of several nearby lakes to operate as micro-hydro PHS facilities. The study focuses on Mountain Lake as the primary upper reservoir (UR), situated at an elevation of approximately 258 meters above sea level. Several surrounding lakes, illustrated in Fig. 6, are evaluated for their potential to serve as lower reservoirs based on their elevation, distance from Mountain Lake, and head difference.

Table 1 summarizes the key physical characteristics of Mountain Lake and its neighboring lakes, detailing elevation above sea level, head difference, shortest horizontal distance to Mountain Lake, and surface area. Notably, Ives Lake, Rush Lake, Pine Lake, Howe Lake, and Lake Superior offer varying elevation differences, with head values ranging from 26.76 to 74.65 meters. However, the significant horizontal distances between Mountain Lake and these lower-lying lakes have prevented conventional hydropower applications. Notably, if the distance between two reservoirs is 1*km* or more, it is more fit for a micro-hydro system, which makes the CPMHS a more practical solution for such cases [2].

The proposed methodology addresses this spatial challenge by implementing a cascade of micro-hydro facilities with intermediate small reservoirs, enabling the utilization of these water resources to generate sustainable power. Fig. 6 provides a 2D schematic diagram of Mountain Lake and its surrounding potential reservoirs, visually supporting the data presented in Table 1. It also demonstrates the feasibility of a stepped, cascading arrangement for energy generation. This approach highlights a potential model for utilizing underexplored water resources for renewable energy, with implications beyond this case study to other lakes in Michigan and across the U.S.

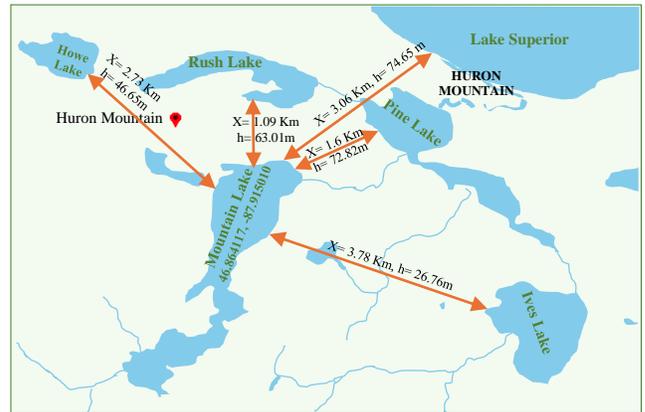

Fig. 6. 2D Schematic Diagram of Mountain Lake and Surrounding Potential Reservoirs.

Table 1: Physical Characteristics of Mountain Lake and Surrounding Lakes for Potential Micro-Hydro Applications.

| Lake Name | Elevation Above Sea Level (m) | Head Difference from Mountain Lake (m) | Shortest Horizontal Distance to Mountain Lake (km) | Surface Area (km²) |
|---|---|---|---|---|
| Mountain Lake | 258.17 | - | - | 3.7439 |
| Ives Lake | 231.41 | 26.76 | 3.78 | 2.0000 |
| Rush Lake | 195.16 | 63.01 | 1.09 | 1.1206 |
| Pine Lake | 185.3 | 72.82 | 1.6 | 3.6653 |
| Howe Lake | 211.52 | 46.65 | 2.73 | 0.7629 |
| Lake Superior | 183.52 | 74.65 | 3.06 | >86,000 |

Fig. 7 illustrates a proposed micro-hydro energy system with Mountain Lake as the primary UR. From this UR, an estimated two surrounding lower reservoirs, Rush Lake and Ives Lake were chosen as a practical application of the proposed methodology. A future optimized system based on the proposed methodology will determine the specific optimal number of those intermediate reservoirs based on optimization and energy management processes. In the left cascade, two intermediate reservoirs lead to Rush Lake (the first lower reservoir), while in the right cascade, three intermediate reservoirs lead to Ives Lake (the second lower reservoir). This estimation is based on the geographical terrain. Each segment's horizontal distance and elevation drop are indicated in Fig. 7, demonstrating the feasibility of energy generation through this stepped, cascading arrangement. This system exemplifies how elevation changes can be leveraged at multiple stages to generate micro-hydro power. By dividing the total elevation drop into manageable segments, each reservoir stage can independently contribute to energy production, collectively enhancing the system's overall energy output. The horizontal distances between each reservoir highlight the innovative use of intermediate reservoirs to bridge otherwise non-viable gaps, transforming challenging terrain into productive micro-hydro sites.

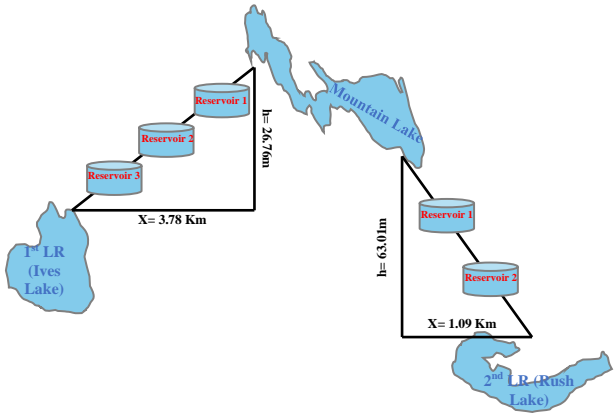

Fig. 7. Conceptual Schematic of Cascading Micro-Hydro System Using Mountain Lake as the Upper Reservoir.

Table 2 provides storage capacities and estimated potential energy outputs for Mountain Lake and its surrounding lakes, offering insight into the system's energy potential. Mountain Lake has a storage volume of 11.23 million $m^3$, serving as the primary energy source for downstream reservoirs. Without a cascading system, the long horizontal distances to Rush Lake and Ives Lake (1.09 km and 3.78 km, respectively) would typically limit energy transfer, making conventional energy extraction less viable.

Table 2: Storage Capacities and Energy Estimates for Mountain Lake and Surrounding Lakes

| Attribute / Lake Name | Mountain Lake | Ives Lake | Rush Lake |
|---|---|---|---|
| Storage volume ($m^3$) | 11.2316 | 3.000 | 3.922 |
| Estimated Storage Energy from Mountain Lake (GWh) | - | 1.1171 | 2.0353 |

However, the proposed cascade configuration significantly enhances energy potential by breaking the elevation drop into smaller stages. As shown in Table 2, the estimated potential energy output for Ives Lake is 1.12 GWh, and for Rush Lake, it is 2.04 GWh, both derived from the head difference relative to Mountain Lake. With a cascading setup, intermediate reservoirs could generate additional energy at each stage, allowing for a more efficient utilization of available head and maximizing the total energy production across the entire system. Such a cascading system will increase the energy generation potential and provide a model for sustainable energy storage in regions with similar topography.

## V. CONCLUSION

This paper presents a robust methodology that addresses the engineering challenge of harnessing micro-hydro storage in regions where significant elevation differences exist, yet the horizontal distance between primary reservoirs is too vast for conventional hydropower solutions. By implementing a system of intermediate cascade reservoirs, this approach enables effective energy transfer and storage, transforming otherwise non-viable sites into productive energy resources. For future work, a comprehensive optimization and energy management strategy will be developed, including realistic data across various scenarios to refine the system's efficiency. This will include full mathematical modeling to evaluate the method's applicability and scalability under diverse conditions, further enhancing its potential for sustainable energy solutions and water management objectives.

ACKNOWLEDGMENT

This work was supported in part by the National Science Foundation of USA under Grant ECCS-2146615 and partially supported by the Department of Energy, Solar Energy Technologies Office (SETO) Renewables Advancing Community Energy Resilience (RACER) program under Award Number DE-EE0010413. Any opinions, findings, conclusions, or recommendations expressed in this material are those of the authors and do not necessarily reflect the views of the Department of Energy.